\documentclass[3p,times]{elsarticle}

\usepackage{ecrc}

\volume{00}

\firstpage{1}

\journalname{Nuclear Physics A}

\usepackage{amssymb}
\biboptions{square,comma,numbers,sort&compress}

\usepackage[figuresright]{rotating}

\usepackage{graphicx}% Include figure files
\usepackage{dcolumn}% Align table columns on decimal point
\usepackage{bm}% bold math
\usepackage{amssymb}
\usepackage{xspace}

\usepackage{hyperref}
\usepackage{amsmath}

\usepackage{color}

\newcommand{\pt}{\ensuremath{p_{\rm{T}}}\xspace}
\newcommand{\pp}{\ensuremath{\rm pp}\xspace}
\newcommand{\ppb}{p-Pb\xspace}
\newcommand{\pbpb}{Pb-Pb\xspace}
\newcommand{\nch}{\ensuremath{N_{\rm ch}}\xspace}

\newcommand{\mpt}{\ensuremath{\langle p_{\rm T} \rangle}\xspace}

\begin{document}
\begin{frontmatter}
%% Title, authors and addresses

%% use the tnoteref command within \title for footnotes;
%% use the tnotetext command for the associated footnote;
%% use the fnref command within \author or \address for footnotes;
%% use the fntext command for the associated footnote;
%% use the corref command within \author for corresponding author footnotes;
%% use the cortext command for the associated footnote;
%% use the ead command for the email address,
%% and the form \ead[url] for the home page:
%%
%% \title{Title\tnoteref{label1}}
%% \tnotetext[label1]{}
%% \author{Name\corref{cor1}\fnref{label2}}
%% \ead{email address}
%% \ead[url]{home page}
%% \fntext[label2]{}
%% \cortext[cor1]{}
%% \address{Address\fnref{label3}}
%% \fntext[label3]{}

\dochead{}

\title{ Mean \pt scaling with $m/n_{\rm q}$ at the LHC: \\ Absence of (hydro) flow in small systems?  }
\author{Antonio Ortiz Vel\'asquez}
%\author[label1]{Antonio Ortiz Velasquez}
%\ead[label1]{antonio.ortiz@nucleares.unam.mx}

\address{Instituto de Ciencias Nucleares, Universidad Nacional Aut\'onoma de M\'exico. \\ Circuito exterior s/n, Ciudad Universitaria, Del. Coyoac\'an, C.P. 04510, M\'exico DF.}

\begin{abstract}

In this work, a study of the average transverse momentum (\pt) as a function of the mid-rapidity charged hadron multiplicity (\nch) and hadron mass ($m$) in \ppb and \pbpb collisions at LHC energies is presented. For the events producing low $\nch$, the average \pt is found to scale with the reduced hadron mass, i.e., mass divided by the number of quark constituents ($m/n_{\rm q}$), this scaling also holds for inelastic \pp collisions at RHIC and LHC energies.  The scaling is broken in high multiplicity \ppb and \pbpb collisions, where, for $\langle {\rm d}N_{\rm ch}/{\rm d}\eta \rangle \lesssim60$ the average \pt is higher for baryons than that for mesons, though they increase linearly with $m/n_{\rm q}$. This behavior is qualitatively well reproduced by Pythia 8, but not by hydro calculations, where an universal scaling with the hadron mass (and not with $m/n_{\rm q}$) is predicted for all the multiplicity event classes. Only the central (0-60\%) \pbpb collisions behave as expected from hydro.

\end{abstract}

\begin{keyword}
%% keywords here, in the form: keyword \sep keyword
Multi-parton interactions, color reconnection, flow-like behavior, nucleon-nucleon reactions, nucleon-nucleus collisions.

%% PACS codes here, in the form: \PACS code \sep code

%% MSC codes here, in the form: \MSC code \sep code
%% or \MSC[2008] code \sep code (2000 is the default)

\end{keyword}

\end{frontmatter}

\section{Introduction}
\label{intro}

The main goal of the heavy ion program is the study of matter under  extreme conditions of high energy density, where QCD phase transitions characterized by deconfinement and chiral symmetry restoration (QGP), are expected to occur~\cite{Adams:2005dq,Adcox:2004mh}. To extract the genuine properties of the hot matter,  the program includes studies in control experiments, like nucleon-nucleon and nucleon-nucleus collisions,  where final state effects were not supposed to be present. Unexpectedly,  recent results at the LHC reveal the presence of QGP-like effects, namely, long range angular correlations~\cite{CMS:2012qk}  and flow-like patterns~\cite{OrtizVelasquez2014146} in the systems created in high multiplicity \pp and \ppb collisions at $\sqrt{s}=7$\,TeV and $\sqrt{s_{\rm NN}}=5.02$\,TeV, respectively.

Such features of data are qualitatively described by hydro calculations~\cite{PhysRevC.88.014903,PhysRevLett.111.172303}, where the formation of a medium and its subsequent fast thermalization are implicitly asummed, but not fully justified. For example, it is known that the validity of a hydrodynamic description requires small  values of Knudsen number ($\ll1$), however, for the systems created in proton-nucleus collisions at $\sqrt{s_{\rm NN}}=5.02$\,TeV, the Knudsen number is already close to one in almost all the space-time points for temperatures $T\lesssim165$\,MeV~\cite{Shen:2015qba}.  Albeit, based on ideas of string percolation~\cite{Gutay:2015cba,Bautista:2015zqu} it has been argued that deconfinement can also occur in small systems, the origin of the phenomenon has not been established~\cite{Antinori:2014xma}. With the same level of accuracy as hydro, other calculations, which do not require the formation of a mini QGP, also do a good job to describe the data. For instance, color glass condensate (CGC), the QCD formulation on the limit of high occupancy number at low $x$-Bjorken, predicts the long range angular correlations and their azimuthal anisotropies for even harmonics\footnote{The odd harmonics were not generated because rescattering  contributions  to  the  intrinsic  correlations  were not  included.} and for \pt $>1$\,GeV/$c$; in \pp and \ppb collisions~\cite{Dumitru:2010iy}. Recently, it has been demonstrated that anisotropic flow ($v_{2}$ and $v_{3}$) can be obtained using gluon distributions from classical Yang-Mills simulations of \ppb collisions~\cite{Schenke:2015aqa}.
Similarly, the AMPT model reproduces the ridge structure in small systems~\cite{Ma:2014pva} and Pythia 8~\cite{Sjostrand:2007gs} with the tune 4C~\cite{Corke:2010yf} is able to generate flow-like patterns using color reconnection (CR) and multi-parton interactions (MPI)~\cite{Ortiz:2013yxa}.  It is worth to mention that early LHC results for \pp collisions, e.g., the transverse sphericity~\cite{Abelev:2012sk} and the di-hadron correlations~\cite{Abelev:2013sqa} as a function of  multiplicity, revealed the importance of multi-parton interactions to describe the data.

As discussed here~\cite{Antinori:2014xma}, it is important to look at new observables which allow one to establish the origin of the QGP-like effects in small systems. In this context, the present work shows a study of the average transverse momentum (\pt) as a function of the hadron mass ($m$) and the event multiplicity (\nch) obtained from different measurements reported by experiments at the LHC. Similarities among \pp, \ppb (all multiplicity classes) and peripheral \pbpb (60-90\%) collisions are found, those effects are qualitatively well captured by Pythia 8, but not by hydro calculations. On the other side, for central Pb-Pb collisions the data exhibit a different feature. 

\section{Multiplicity dependence of the average \pt vs. hadron mass in Pythia}

\begin{figure}
\begin{center}
\resizebox{0.75\textwidth}{!}{%
  \includegraphics{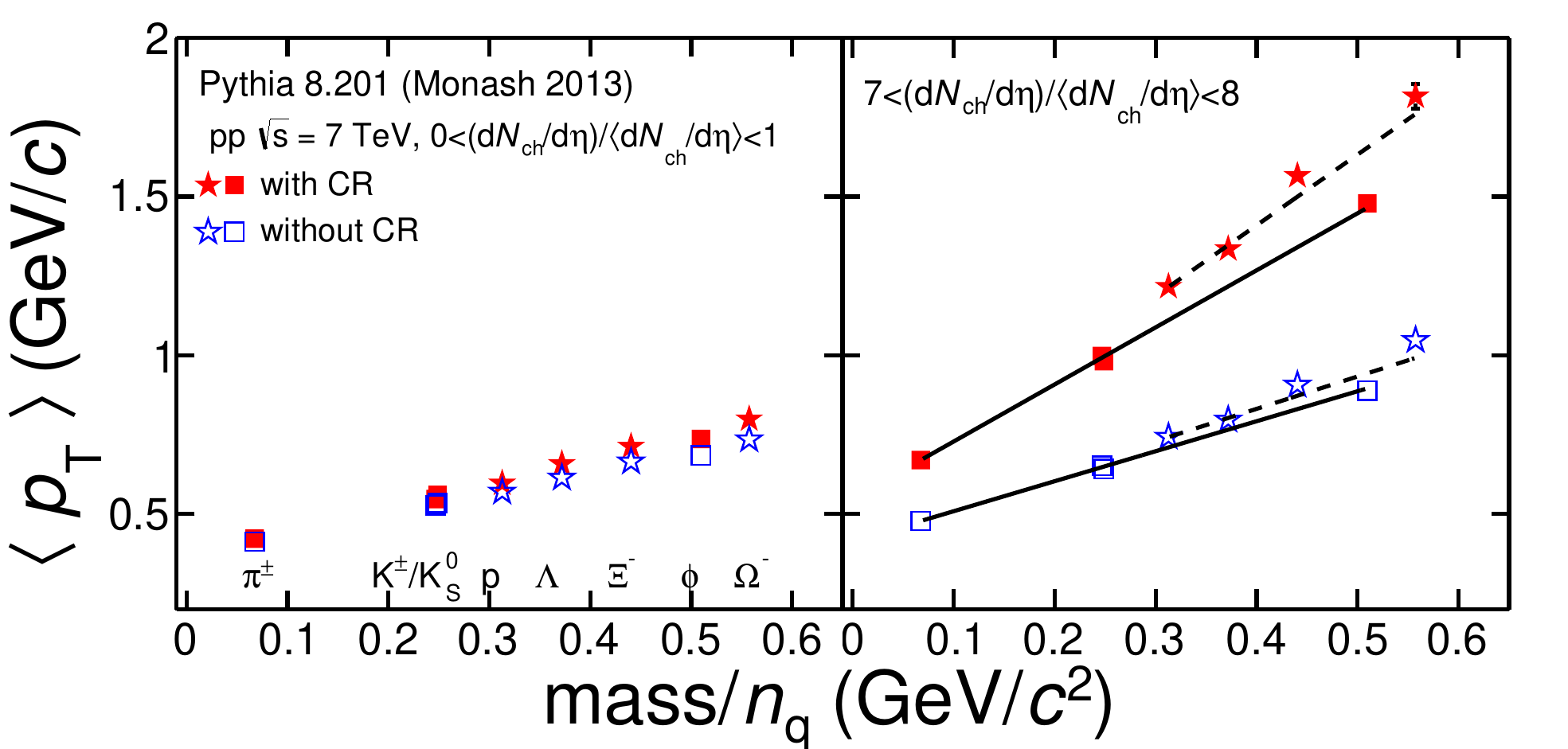}
}
\caption{(Color online). Average transverse momentum as a function of the scaled hadron mass ($m/n_{\rm q}$) for pp collisions at $\sqrt{s}=7$\,TeV simulated with Pythia 8.201. Two multiplicity event classes are shown, low multiplicity (left) and high multiplicity (right). Results are presented for simulations with (full red markers) and without (empty blue markers) color reconnection. The solid (dashed) line indicates a linear fit to meson (baryon) data.}
\label{fig:2}       % Give a unique label
\end{center}
\end{figure} 

In Pythia 8, color reconnection and multi-parton interactions produce flow-like patterns in \pp collisions at the LHC energies via the creation of boosted strings~\cite{Ortiz:2013yxa}.  To illustrate this, the MC \pt spectra of different particle species ($\pi^{+}+\pi^{-}$, ${\rm K}^{+}+{\rm K}^{-}$,  ${\rm K_{S}^{0}}$, ${\rm p+\bar{p}}$, $\phi$-meson, $\Lambda+\bar{\Lambda}$, $\Xi^{-}+\bar{\Xi}^{+}$ and $\Omega^{-}+\bar{\Omega}^{+}$)  for \pp collisions at  $\sqrt{s}=7$\,TeV were studied using Pythia 8.201~\cite{Sjostrand:2014zea} (tune Monash 2013~\cite{Skands:2014pea}). The transverse momentum spectra and event multiplicity were determined at $|y|<1$ and $|\eta|<1$, respectively.  It has been shown that the Boltzmann-Gibbs blast-wave function~\cite{PhysRevC.48.2462}  describes within 10\% the \pt distributions of all hadrons under consideration when color reconnection is activated~\cite{Cuautle:2015kra}. Actually, the ALICE Collaboration has reported the same level of agreement (within $\approx$10\%) between the data and the blast-wave model for \ppb collisions~\cite{Abelev:2013bla}. Furthermore, with CR the correlation between the parameters obtained from the fit to MC \pt spectra, the average transverse velocity, $\langle \beta_{\rm T} \rangle$, and the temperature at the kinetic freeze-out, $T_{\rm kin}$; behaves like in data~\cite{Abelev:2013bla, Cuautle:2015kra}. Namely, $\langle \beta_{\rm T} \rangle$ increases with increasing the multiplicity density (${\rm d}N_{\rm ch}/{\rm d}\eta$), and at the same time $T_{\rm kin}$ decreases\footnote{In simulations without color reconnection, the multiplicity dependence is very weak, i.e., $\langle \beta_{\rm T} \rangle$ and  $T_{\rm kin}$  do not evolve with multiplicity.}. Hence, the results from the blast-wave analysis are not enough to draw a final conclusion on the possible formation of a strongly interacting QCD medium in high multiplicity \pp and \ppb collisions. 

\begin{figure}
\begin{center}
\resizebox{1.05\textwidth}{!}{%
  \includegraphics{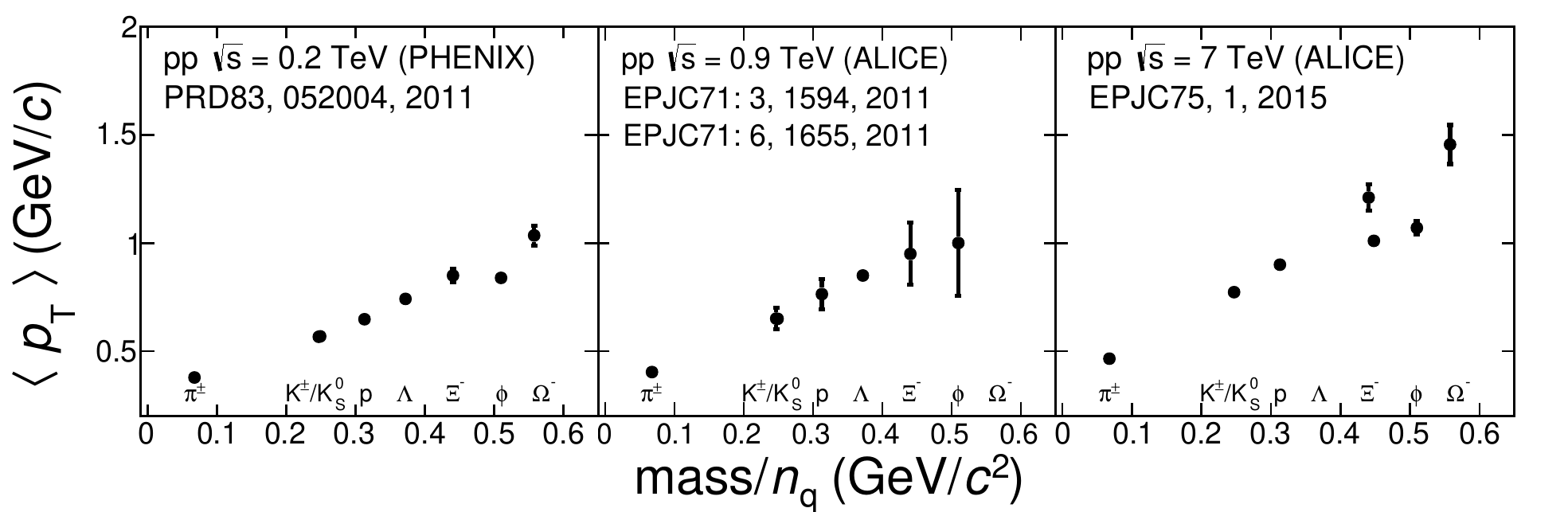}
}
\caption{Average transverse momentum as a function of the scaled hadron mass ($m/n_{\rm q}$) for minimim bias \pp collisions at $\sqrt{s}=0.2$\,TeV (left), 0.9\,TeV (middle) and  7\,TeV (right) measured by PHENIX and ALICE experiments.}
\label{fig:2a}       % Give a unique label
\end{center}
\end{figure}

The main focus of this work is the study of the mass dependence of the average \pt. In the string hadronization model, the string boost is transferred to the final particles when the string breaks up and for the same boost velocity a heavy particle will gain more \pt than a lighter one. CR is therefore expected to give a  rise of the average \pt vs. hadron mass. Qualitatively, the effect is in good agreement with data, but so far, the current models of CR are not able to reproduce the size of the effect~\cite{Christiansen:2015yqa}. However, other effects look promising, for example the  color ropes formation, which by increasing the effective string tension, can also produce flow-like behavior~\cite{Bierlich:2014xba}.  

The mass dependence of the average \pt has been investigated for two cases, with and without color reconnection. When color reconnection is deactivated, the partonic systems fragment independently, and even in this case, the average \pt increases with the hadron mass. It is worth to remember that in LHC data, even for the lowest multiplicity event class, where a medium is not expected to be formed, the mass ordering is also seen~\cite{Abelev:2013haa,Chatrchyan:2013eya,Chatrchyan:2012qb}. The \mpt increase with the hadron mass, therefore is not a distinct feature of hydro, instead it seems to be a characteristic of the fragmentation. Actually, another observation which is reported for the first time is that Pythia 8.201 without CR produces an average \pt  which scales with the reduced hadron mass, i.e., mass divided by the number of quark constituents ($m/n_{\rm q}$), and not with the hadron mass: 

\begin{equation}
\mpt \propto \frac{m}{n_{\rm q}}.
\end{equation}

This is illustrated in Fig.~\ref{fig:2}, where the correlation between \mpt and $m/n_{\rm q}$ is shown for two event classes: low ($0 < {\rm d}N_{\rm ch}/{\rm d}\eta <   \langle {\rm d}N_{\rm ch}/{\rm d}\eta \rangle$) and high ($ 7\times\langle {\rm d}N_{\rm ch}/{\rm d}\eta \rangle < {\rm d}N_{\rm ch}/{\rm d}\eta <   8\times\langle {\rm d}N_{\rm ch}/{\rm d}\eta \rangle$) multiplicity. The scaling is valid across the multiplicity event classes, though, it is somewhat violated  at high multiplicity.  Figure~\ref{fig:2} also illustrates the results when color reconnection is activated, in this case, at low multiplicity the scaling holds, while at high \nch it is broken and the \mpt for baryons and mesons increase with different slopes. In addition, the mean \pt for $\phi$ meson is higher to that for protons, hence, neither a scaling with hadron mass would be expected.  

In the hadronization model used in Pythia, the constituent quark number dependence can be built up as follows. Quark - (anti)quark pairs are produced during the string break-up process, a transverse momentum is assigned to the pair following a flavor-independent Gaussian spectrum. Since no string transverse excitations are assumed, the \pt is locally compensated between the quark and (anti)quark of the pair. Each hadron is subsequently formed with the string segments and the total partonic \pt is assigned to the hadron. In this model, a meson corresponds to a small piece of string with a quark in one end and an (anti)quark in the other~\cite{Sjostrand:2006za}. This fragmentation is universal, in the sense that is the same for all the primary hadrons produced in the collisions, and therefore a dependence of the average hadron \pt with the constituent quark number is expected to occur. Following the same prescription, in events with large number of MPI and with CR, partons from different origin (MPI, initial state radiation, etc.) are allowed to recombine to form hadrons. This sort of recombination leads to a general $n_{\rm q}$ dependence of the \pt which becomes more complex since, as discussed in the previous paragraph, on top of the universal Gaussian fragmentation the string segments are also boosted. In contrast, hydro calculations for small systems~\cite{Bozek:2014era} give a \mpt which is independent of $n_{\rm q}$.

The next sections are dedicated to study the same effects in different colliding systems using published LHC data.

\section{Study of  \mpt using LHC data}

The scaling of \mpt with $m/n_{\rm q}$ is first investigated in inelastic \pp collisions at different center of mass energies, namely, 0.2\,TeV, 0.9\,TeV and 7\,TeV. Data were obtained from different publications of experiments at the RHIC~\cite{Adare:2010fe} and at the LHC~\cite{Aamodt:2011zza,Aamodt:2011zj,Abelev:2014qqa}. Figure~\ref{fig:2a} shows the correlation plots, \mpt vs. $m/n_{\rm q}$, which are obtained. For 0.2\,TeV data, where the event multiplicity is the smallest, $\langle {\rm d}N_{\rm ch}/{\rm d}\eta \rangle \approx$ 2.25~\cite{Alver:2010ck}, the $m/n_{\rm q}$ scaling scaling holds even if heavy hadrons like $\Xi^{-}$ and $\Omega^{-}$ are considered.  The same observation is valid for \pp collisions at 0.9\,TeV, where the event multiplicity is still small, $\langle {\rm d}N_{\rm ch}/{\rm d}\eta \rangle \approx$ 3.81~\cite{Aamodt:2010pp}.  For the highest center of mass energy, where the identified hadron \pt spectra are available, the scaling is slightly broken as observed in Pythia 8.201 with color reconnection. As in the model, the \mpt for baryons rises faster with the reduced mass than for mesons. Also the average \pt for baryons is higher to that for mesons.

\begin{figure*}
\begin{center}
\resizebox{0.75\textwidth}{!}{%
  \includegraphics{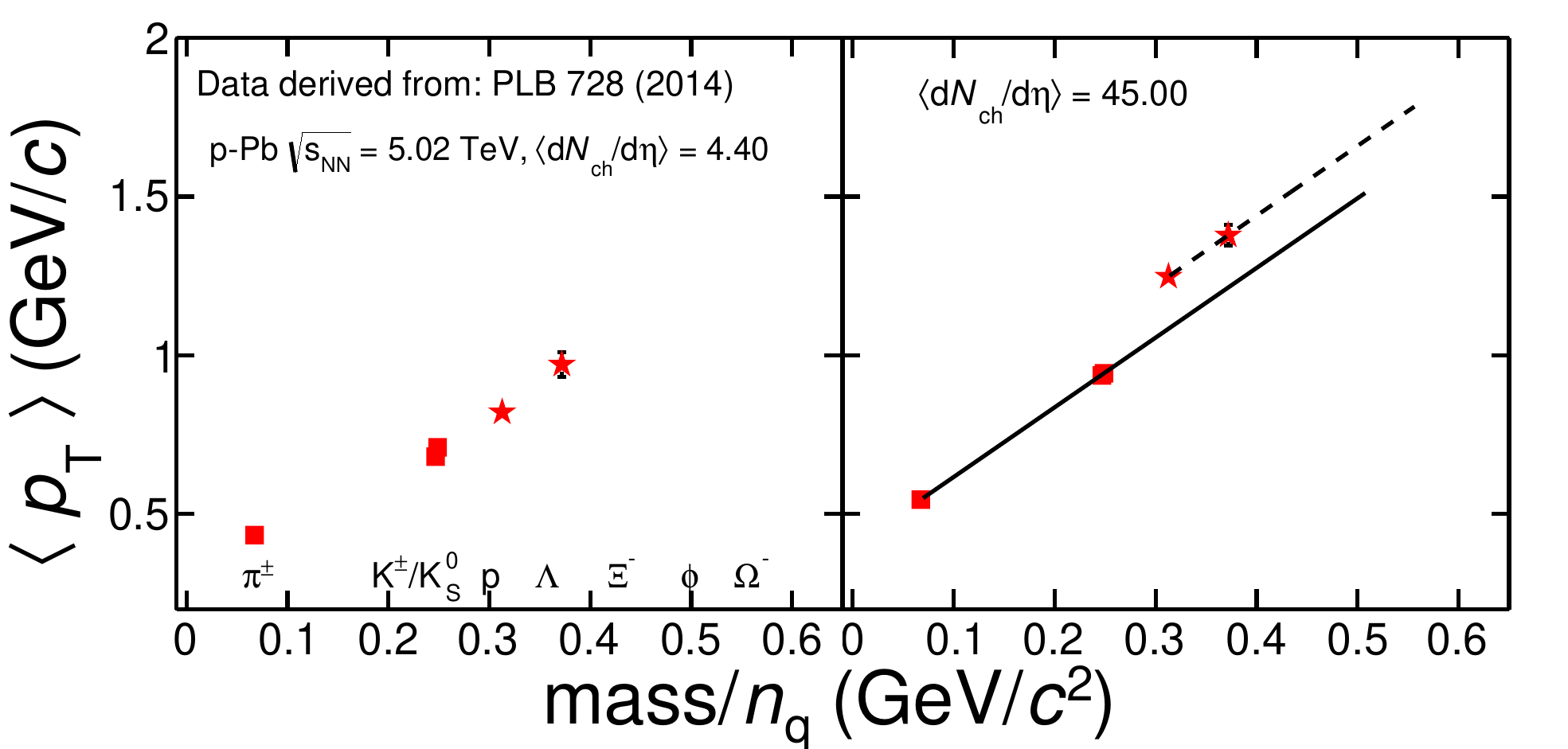}
}
\resizebox{0.75\textwidth}{!}{%
  \includegraphics{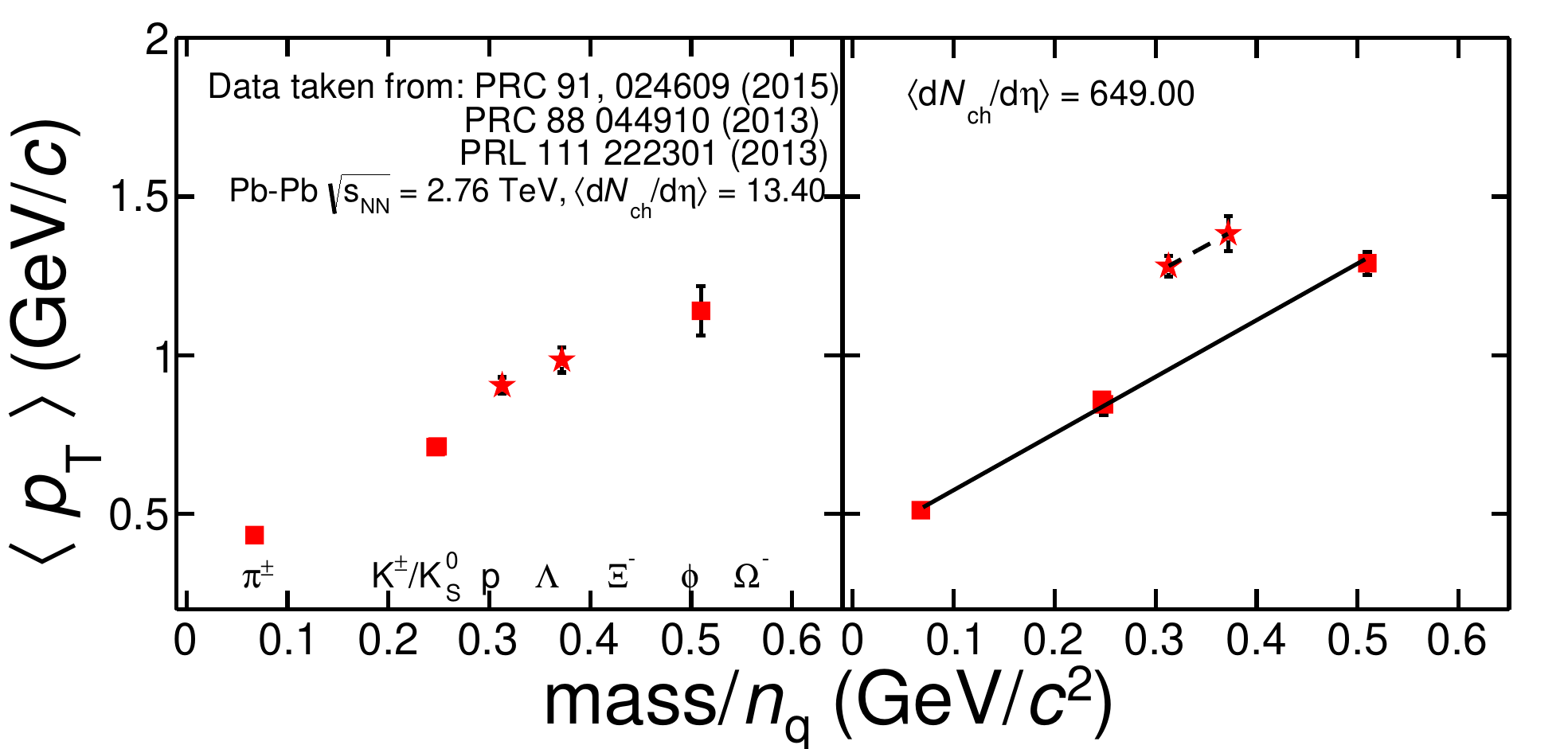}
}
\vspace*{0.5cm}       % Give the correct figure height in cm
\caption{(Color online). Average transverse momentum as a function of the hadron mass normalized to the number of constituent quarks. Results for the lowest and highest multiplicity event classes in \ppb collisions are shown in the upper panel. For \pbpb collisions, the results for the most peripheral and the most central events are shown in the bottom panel. The solid (dashed) line indicates a linear fit to meson (baryon) data. The average multiplicity densities, $\langle {\rm d}N_{\rm ch}/{\rm d}\eta \rangle$, for the different event classes have been measured by ALICE in $|\eta|<0.5$~\cite{Abelev:2013haa,Aamodt:2010cz}.}
\label{fig:3}       % Give a unique label
\end{center}
\end{figure*}

Using the existing ALICE data for \ppb collisions~\cite{Abelev:2013haa}, the average \pt for $\pi^{+}+\pi^{-}$, ${\rm K}^{+}+{\rm K}^{-}$, ${\rm p+\bar{p}}$, ${\rm K}_{S}^{0}$ and $\Lambda+\bar{\Lambda}$ as a function of $m/n_{\rm q}$ are obtained and plotted in the upper part of Fig.~\ref{fig:3}. Two multiplicity event classes are shown, $\langle {\rm d}N_{\rm ch}/{\rm d}\eta \rangle$=4.4 and $\langle {\rm d}N_{\rm ch}/{\rm d}\eta \rangle$=45. As in Pythia 8, for low multiplicity events the scaling of \mpt with $m/n_{\rm q}$ is observed, while for high multiplicity events it is broken. There, the mean \pt for baryons is higher than for mesons. The analogous analysis for \pbpb data~\cite{Abelev:2014uua,Abelev:2014laa,Abelev:2013xaa} is presented in the bottom panel of Fig.~\ref{fig:3}, there, also the $\phi$ meson \mpt is reported. Like in Pythia 8, the scaling of \mpt with $m/n_{\rm q}$ is only valid for the most peripheral \pbpb collisions (80-90\%). For the most central \pbpb collisions (0-5\%), the mean \pt for baryons and mesons as a function of  $m/n_{\rm q}$ exhibit different trends. In addition, for this event class the average \pt for protons is roughly the same to that for $\phi$ mesons. This observation triggers the question of whether a scaling of \mpt with mass, and not with $m/n_{\rm q}$, is seen in the most central \pbpb events. 

\begin{figure}
\begin{center}
\resizebox{1.05\textwidth}{!}{%
  \includegraphics{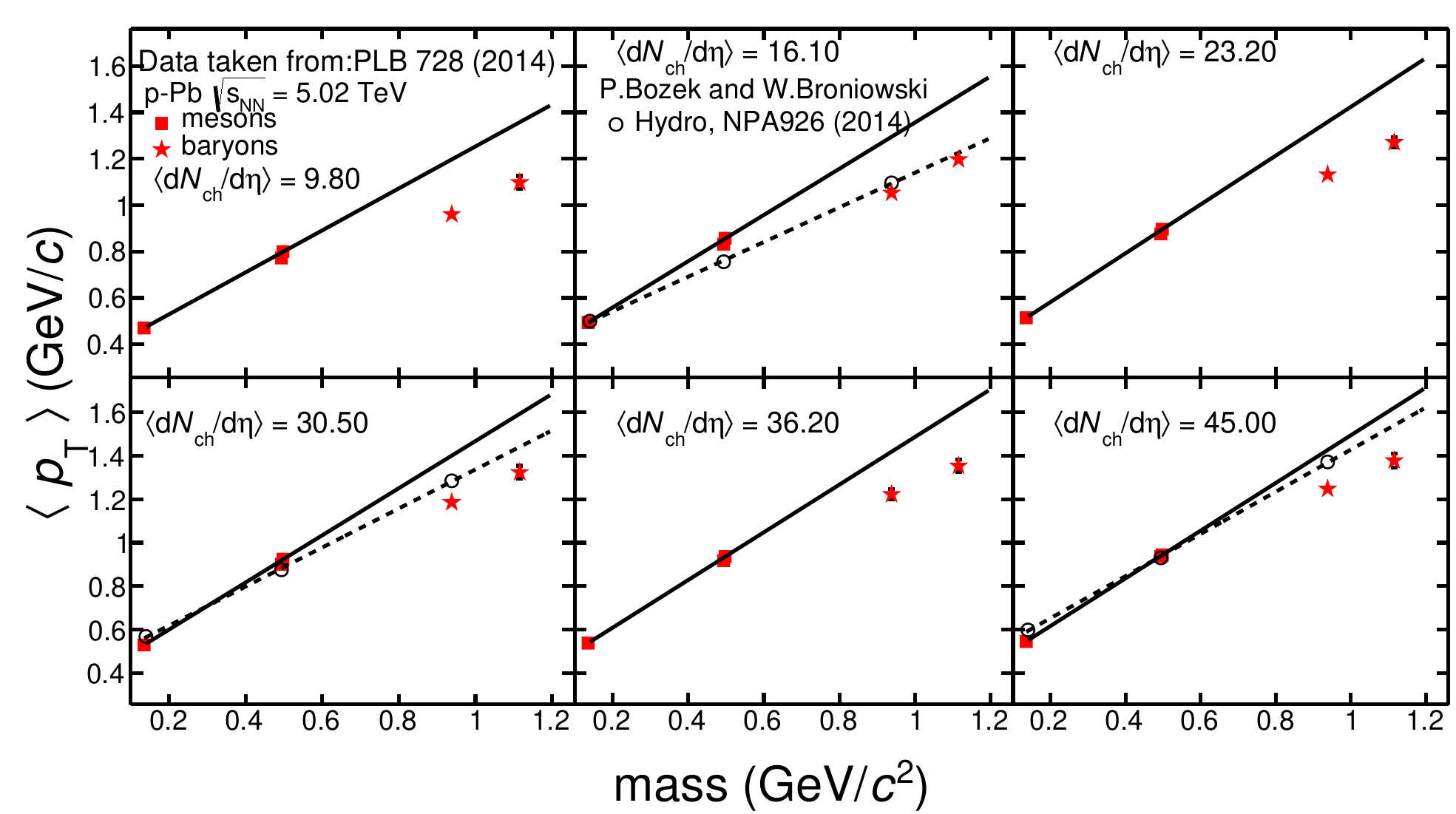}
}
\caption{(Color online). Multiplicity dependence of the average transverse momentum as a function of the hadron mass in \ppb collisions at $\sqrt{s_{\rm NN}}=5.02$\,TeV (solid markers).  A first degree polynomial (solid line) is fitted to the meson data. For three multiplicity classes, hydro calculations~\cite{Bozek:2014era} are also shown (empty markers), linear fits to meson \mpt calculations are shown as dashed lines. The average multiplicity densities, $\langle {\rm d}N_{\rm ch}/{\rm d}\eta \rangle$, for the different event classes have been measured by ALICE in $|\eta|<0.5$~\cite{Abelev:2013haa}.}
\label{fig:5}       % Give a unique label
\end{center}
\end{figure}

\begin{figure}
\begin{center}
\resizebox{1.05\textwidth}{!}{%
  \includegraphics{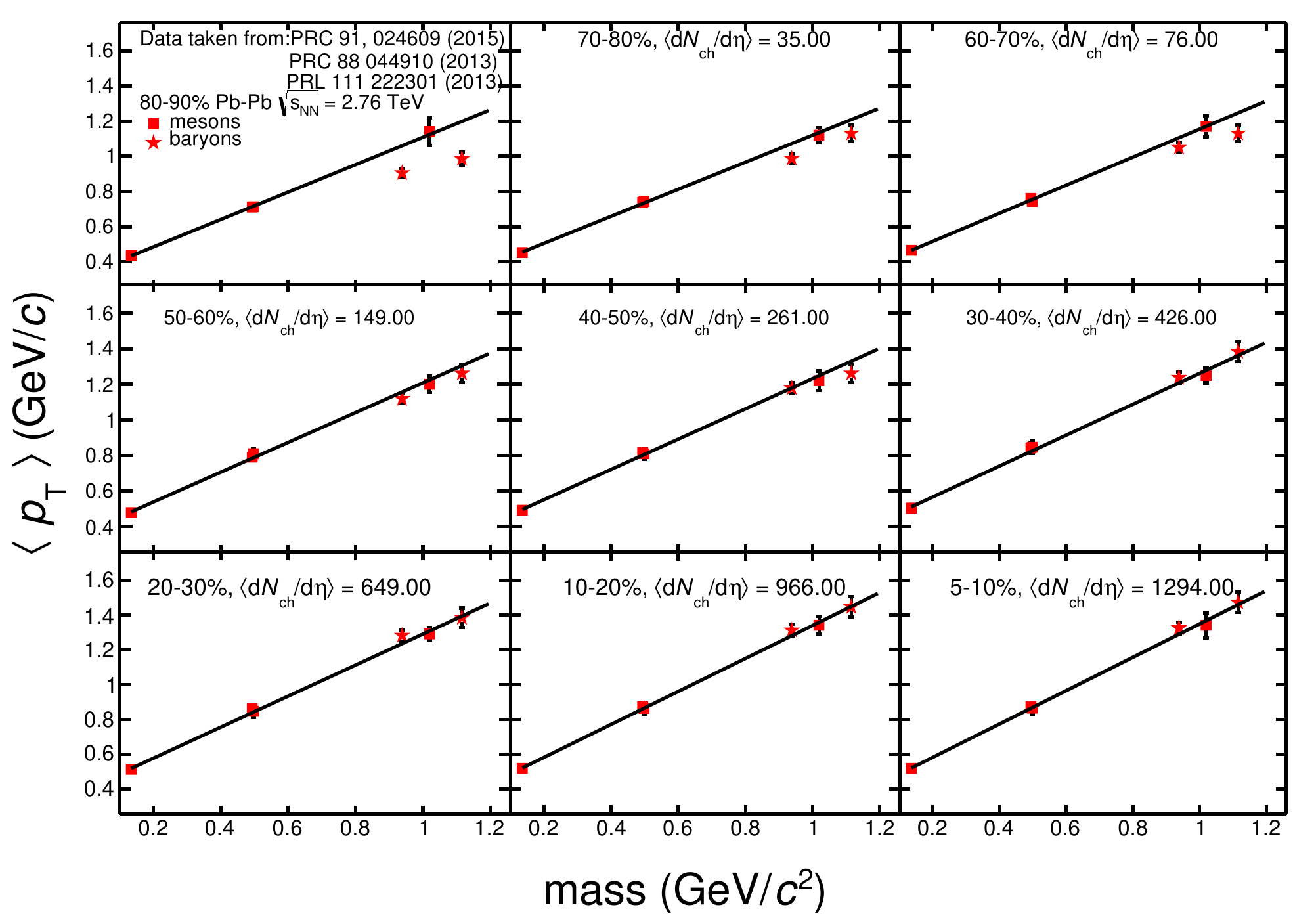}
}
\caption{(Color online). Centrality dependence of the average transverse momentum as a function of the hadron mass in \pbpb collisions at $\sqrt{s_{\rm NN}}=2.76$\,TeV. Solid line indicates the first degree polynomial which fits the meson data. The average multiplicity densities, $\langle {\rm d}N_{\rm ch}/{\rm d}\eta \rangle$, for the different event classes have been measured by ALICE in $|\eta|<0.5$~\cite{Aamodt:2010cz}.}
\label{fig:6}       % Give a unique label
\end{center}
\end{figure}

The possible universal scaling with $m$ is investigated, Figs.~\ref{fig:5} and  ~\ref{fig:6} show the multiplicity dependence of the average \pt as a function of the hadron mass for \ppb and \pbpb collisions, respectively. The universal mass scaling is only observed in the 0-60\% centrality classes for \pbpb collisions. While, it is broken for the rest of the \pbpb centrality classes (60-90\%) and for all the multiplicity classes measured in \ppb collisions. In Fig.~\ref{fig:5}, also hydro calculations for \ppb~\cite{Bozek:2014era}  are shown, there, the mass scaling holds for all the multiplicity classes.

\begin{figure}
\begin{center}
\resizebox{0.55\textwidth}{!}{%
  \includegraphics{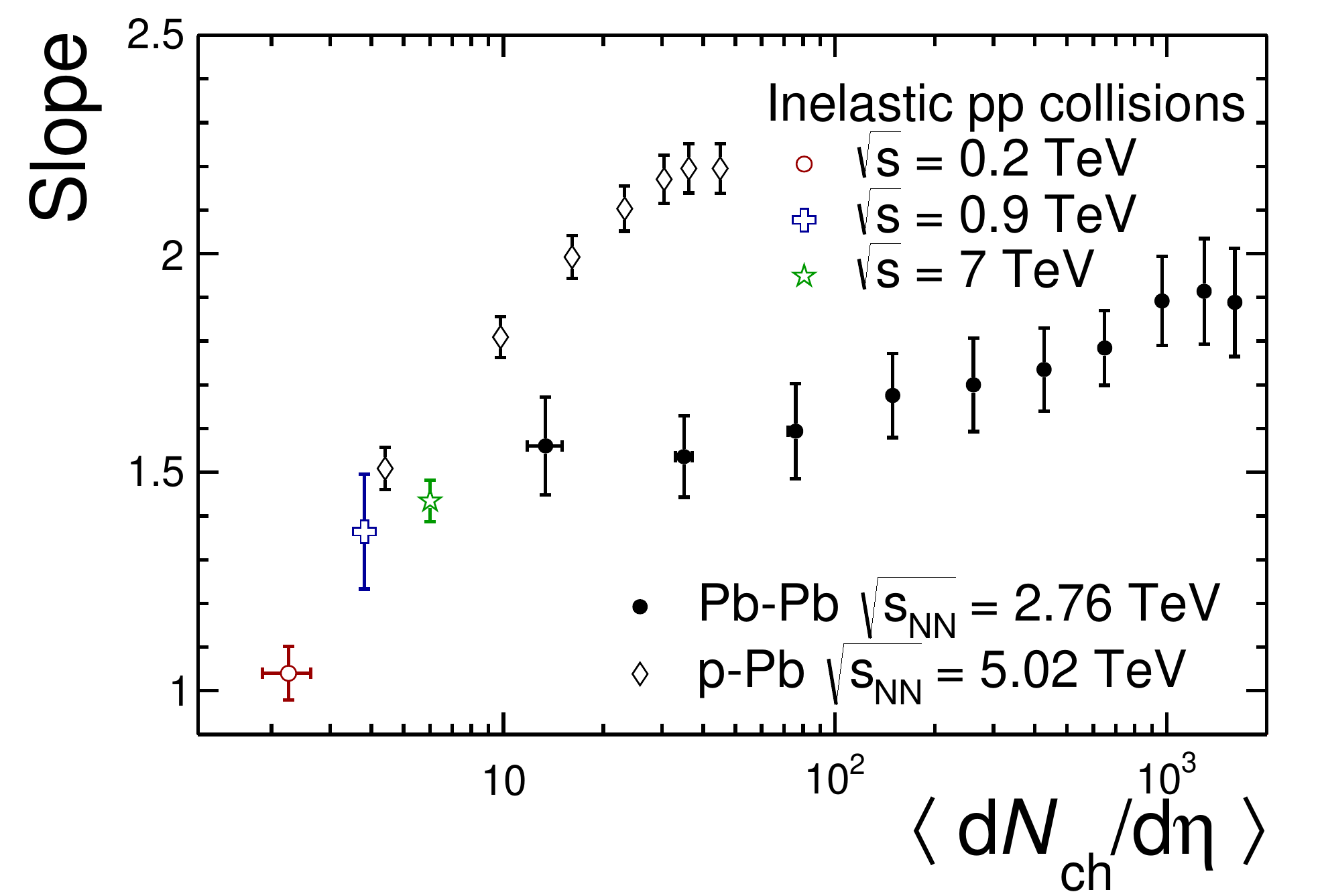}
}
\caption{(Color online). For each event class, a first degree polynomial is fitted to the average \pt vs. the reduced hadron mass ($m/n_{\rm q}$). In this figure, the slopes extracted from the fits are plotted as a function of the event multiplicity for different colliding systems.}
\label{fig:7}       % Give a unique label
\end{center}
\end{figure}

One thing which can be seen is that for the same multiplicity density, $\langle {\rm d}N_{\rm ch}/{\rm d}\eta \rangle \approx$35, the slope of \mpt vs $m$ is higher in \ppb than in \pbpb collisions. This can be quantified if a first degree polynomial is fitted to data, in this exercise the correlations \mpt vs $m/n_{\rm q}$ for \ppb and \pbpb data were used. Figure~\ref{fig:7} show the results for mesons, where the slopes obtained from the fits are plotted as a function of the event multiplicity. For comparison, also the results for inelastic \pp collisions at different center of mass energies are shown. At low multiplicity, where the universal scaling of \mpt vith $m/n_{\rm q}$ is approximately valid, the data points from all the systems follow the same trend, and then, for higher multiplicities, \ppb and \pbpb data deviate.

%\section{Discussion}

\section{Conclusions}

Collectivity is considered one of the key ingredients to establish the formation of an equilibrated QGP. So far, the way to verify whether a system behaves collectively or not relies in the success of the hydro models to describe flow observables. Namely, the \pt distributions of identified hadrons and $v_{\rm n}$, which are connected with  radial and anisotropic flow, respectively.  In the present work the details of existing \pt spectra for different colliding systems were studied aiming to check two things. On one side, how precise hydro models describe data, specifically, to check in data the mass scaling of the average \pt as predicted by hydro~\cite{Bozek:2014era}. And on the other side, to test an alternative explanation based on color reconnection and multi-parton interactions. This investigation brought the following results:
\begin{itemize}
\item Low multiplicity \pp collisions simulated with Pythia 8 exhibit an universal scaling of \mpt with $m/n_{\rm q}$ (mass divided by the number of quark constituents). When CR is switched-off, the effect does not depend on multiplicity. In the opposite case, this universal scaling is broken, i.e., a deviation from the linear behavior is observed, and the average \pt for baryons and mesons increase following different lines. 
\item LHC data for \ppb collisions as a function of the event multiplicity give results which are qualitatively similar to those observed in Pythia 8.  The same observation holds for inelastic \pp collisions at RHIC and LHC energies. 
\item For LHC \pbpb data  the average \pt as a function of mass shows an interesting phenomenon. Up to centralities of about 50-60\% the baryon and meson \mpt do not scale with the same dependence (similar to the results obtained with Pythia). For more central collisions the scaling with mass seems to work.
\end{itemize}
The results imply that the application of the hydro model should be done with great care because the scaling with mass is not an universal behavior of the data. The hydro model assumes that the behavior of all particles in the mostly pion cloud created after the collision is identical. However, without taking into account the difference in the pion - particle interaction cross section (as was observed at SPS and RHIC energies for the slope). Again the results of Pythia simulations surprisingly well reproduce the different behavior of mesons and baryons raising the possibility of an altogether alternative interpretation.

\section{Acknowledgments}
The author acknowledges the useful discussions with Peter Christiansen, Eleazar Cuautle,  Torbj{\"o}rn Sj{\"o}strand and  Guy Pai\'c. Support for this work has been received by CONACyT
under the grant No. 260440; and by DGAPA-UNAM under PAPIIT grants IA102515, IN105113, IN107911 and IN108414.

\bibliographystyle{elsarticle-num}
%\bibliography{<your-bib-database>}

\bibliography{biblio}% Produces the bibliography via BibTeX.

\end{document}